\documentclass[10pt]{article}
\usepackage[dvips]{graphicx,psfrag}
\usepackage{amssymb}
\usepackage{color}
\input epsf
\definecolor{Blue}{rgb}{0.3,0.3,0.9}
\definecolor{Red}{rgb}{1,0,0}
\definecolor{Green}{rgb}{0,1,0}
\newcommand{\be}{\begin{equation}}
\newcommand{\ee}{\end{equation}}
\newcommand{\bea}{\begin{eqnarray}}
\newcommand{\eea}{\end{eqnarray}}

\begin{document}

\begin{center}
{\Large\bf World-making with extended gravity black holes for cosmic natural
selection in the multiverse scenario}

\bigskip \bigskip \medskip
{\large Aur\'elien Barrau
}\\[.5cm]
{\it Laboratory for Subatomic Physics and Cosmology, CNRS/IN2P3, Grenoble
Universit\'es\\ 53, avenue de Martyrs, 38026 Grenoble cedex, France }\\[1cm]

{\bf Abstract}
\end{center}
Physics is facing contingency. Not only in facts but also in laws (the frontier becoming extremely narrow). Cosmic natural selection is a tantalizing idea to explain the apparently highly improbable structure of our Universe. In this brief note I will study the creation of Universes by black holes in --string inspired-- higher order curvature gravity.
\\[1cm]


~~~~~~~Keywords~: Anthropic principle, Lovelock black holes, String gravity.\\[1cm]

Cosmology is a very strange branch of physics which contradicts nearly all the 
heuristic principles of the Science of Nature~: the observer is part of the system described, the experiment "creation of the Universe" is unique (or has long been
thought to be), the boundary conditions are relevant for the fundamental laws,
the history must be extrapolated backward in time --as forbidden by the second
law of thermodynamics-- and the energy scales considered are well above all that can
be probed by laboratory experiments. Beyond all such specific features, this
framework makes it unavoidable to question ourselves about the {\it necessity} of things and rules
as they now appear. Among possible values of fundamental parameters, those
leading to the "kind of Universe" we know are often located in a zero measure
phase space. This surprising situation could be naturally understood if there is
not a single Universe but many (maybe an infinity) of them with different
realizations of the science laws, as suggested by the string landscape \cite{susskind}. We simply live in one of them which is
compatible with the existence of our {\it specific} life (the literature on the
anthropic principle is very rich, a good up-to-date summary is given in
\cite{carter} by Brandon Carter who first introduced this concept in modern
cosmology). The idea of multiple Worlds, which is necessary for this viewpoint to be conclusive beyond a simple space selection effect, is far from being new or revolutionary. It was probably introduced
by Anaximander and Democritus in ancient Greece, revisited by Nicholas from Cusa
and Giordano Bruno at the
beginning of Renaissance (together with Rabelais refering to the Pythagorean 
philosopher Petron) before being --with very different meanings--
considered by Leibniz and Spinoza in the XVII$^{{\rm th}}$ century. Since this time, it has
never been abandoned. In contemporary
philosophy, following somehow the "Pluralistic Universe" of William James, it
lies at the basis of many thoughts from Goodman \cite{goodman}, Foresta
\cite{foresta}, Hintikka \cite{hintikka} and Lewis \cite{lewis}, among many others.\\

\begin{figure} 
	\begin{center}
		\includegraphics[scale=0.6]{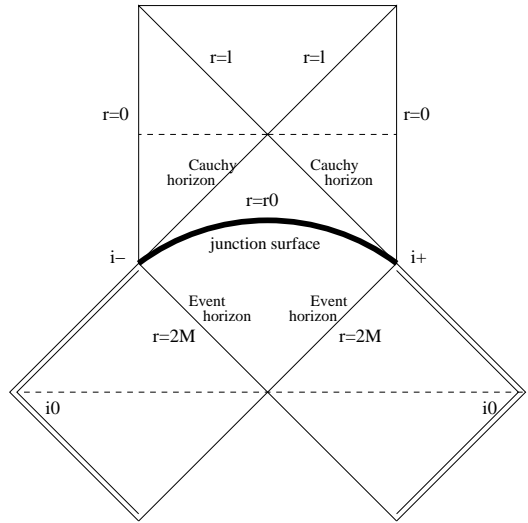}
		\caption{Conformal Penrose-Carter diagram for a Scwarzschild
		black hole with a de Sitter interior \cite{frolov}.}
		\label{fro}
	\end{center}
\end{figure}

As far as physics in concerned, I will focus in the following on the proposal
from Lee Smolin \cite{smolin} (which has already been objected some pertinent arguments 
\cite{vil}) where black holes are considered as creators of
new Universes. This alternative to the more popular view where chaotic inflation
plays this role is motivated by its ability to allow for inherited characters
(what I will call a Lamarck Universe), to rely on better understood physical
grounds and to develop a causal structure for selection effects to occur. In
this scenario, our Universe would have been born from a black hole in another Universe
and would have generated at least $10^{15}$ Universes through the black holes 
it certainly contains (this could have interesting implications on the usual 
question about the surprisingly small value of the entropy of our Universe). 
The main idea relies on bouncing which is quite well established for 
cosmological models but remains quite controversial in the black hole sector. 
Frolov, Markov \& Mukhanov \cite{frolov} have shown how the Schwarzschild metric
inside a black hole can be attached to the de Sitter one at some spacelike
junction surface leading to the spacetime displayed in Fig.~\ref{fro}.

In addition to controversial postulates about the equation of state close to 
the central singularity ($R^2=4R^{\nu}_{\mu}R^{\mu}_{\nu}$), 
the basic assumption is the existence of a limiting curvature~: $R_{\mu\nu\lambda\rho}R^{\mu\nu\lambda\rho}<\alpha/l^4$ where $l$ has the dimension of a length (expected to lie close to the Planck size) and $\alpha\sim 1$.

This {\it ad hoc} assumption can be investigated in the framework of string gravity, which is expected to account for {\it some} quantum effect in the gravitational sector.
Let us consider first the general 4-dimensional effective string action with second order curvature corrections~:

$$S= \frac{1}{16\pi} \int d^4 x \sqrt{-g} \biggl[ - R+ 2 \partial_{\mu} \phi \partial^{\mu}
\phi + \lambda_2 e^{-2\Phi} ( R_{\mu\nu\lambda\rho}R^{\mu\nu\lambda\rho} -4R_{\mu\nu}r^{\mu\nu}+R^2) \biggr].$$

This Lagrangian is well known for appearing in the limit of different string 
theories \cite{mignemi}. It can also be viewed as a second order truncature of Lovelock 
gravity --made dynamical in 4 dimensions through the dilatonic coupling-- which
can be seen as the 
sum of dimensionally extended Euler densities. The full Lagrangian density 
(which corresponds to the most general tensor that can be equaled to the
stress-energy tensor while keeping the mathematical structure of the theory) reads as \cite{myers}
$$L=\Sigma_{n=0}^{k}c_nL_n~{\rm
with}~L_n=2^{-n}\delta_{c_1d_1c_nd_n}^{a_1b_1a_nb_n}
R^{c_1d_1}_{~~~~a_1b_1}...R^{c_nd_n}_{~~~~a_nb_n},$$
where $c_i$ are constants, $\delta$ is totally antisymmetric in both sets of
indices and $L_n$ is the Euler density of a 2n-dimensional manifold.

In any attempt to perturbatively quantize gravity as a field theory, higher-derivative
interactions must be included in the action. Furthermore, higher-derivative gravity theories are 
intrinsically attractive as in many cases
they display features of renormalizability and asymptotic freedom. Among such approaches,
Lovelock gravity \cite{love} is especially interesting as the resulting equations of
motion contain no more than second derivatives of the metric, include the self interaction
of gravitation, and are free
of ghosts when expanding around flat space. The
four-derivative Gauss-Bonnet term is most probably the dominant correction to the 
Einstein-Hilbert
action \cite{zie}. Gauss-Bonnet gravity was shown to exhibit a very
rich phenomenology in cosmology (see, {\it e.g.}, \cite{char} and references
therein), high-energy physics (see, {\it e.g.}, \cite{barr} and references
therein) and black hole theory (see, {\it e.g.}, \cite{alex} and references
therein). It also provides interesting solutions to
the dark energy problem \cite{Nojiri}, offers a promising framework for
inflation \cite{lidsey}, allows useful modification of the Randall-Sundrum model 
\cite{kim} and, of course, solves most divergences associated with the endpoint of
the Hawking evaporation process \cite{alex2}.

Those Gauss-Bonnet black-holes can be studied with the choice of metric~:
$$
ds^2 = \Delta dt^2 - \frac{\sigma^2}{\Delta} dr^2 - f^2
(d \theta^2 + \sin^2 \theta d \varphi^2).
$$
Integrating numerically (using an additional parameter) inside the event 
horizon $r_h$ allows to show \cite{alex} that a solution exists only up to a
 (lower) value of the radius $r=r_s$. A new branch exists between $r_s$ and a 
 singular horizon $r_x$ but its physical meaning is unclear. The main point 
I want to underline is that the curvature invariant can be computed \cite{alex} around the 
 minimal radius $r_s$ with the parameterization $x=\sigma-\sigma_s$~:

\begin{eqnarray*}
R_{\mu\nu\rho\lambda}R^{\mu\nu\rho\lambda} & = &
4 \frac{\Delta^2}{\sigma^4 r^4} +
8 \frac{\Delta^2 (\sigma')^2}{\sigma^6 r^2} -
8 \frac{\Delta}{\sigma^2 r^4} -
8 \frac{\Delta \Delta' \sigma'}{\sigma^5 r^2}
+ \frac{(\Delta'')^2}{\sigma^4} \nonumber \\
& + &
4 \frac{(\Delta')^2}{\sigma^4 r^2} -
2 \frac{\Delta'' \Delta' \sigma'}{\sigma^5} +
\frac{(\Delta')^2 (\sigma')^2}{\sigma^6} +
\frac{4}{r^4} \nonumber \\
& = & \frac{1}{x^2} \ \biggl( \frac{4 d_s}{r_2 r^2_s \sigma^6_s} +
\frac{d_2}{2 r^2_2 \sigma^6_s} \biggr) + O\biggl(\frac{1}{x} \biggr)
\rightarrow \infty
\end{eqnarray*}
where $d_2$ and $d_s$ are expansion coefficients.

This means that {\bf although the correct topology $S^2\times R^1$ (an infinite --in the $t$ direction-- tube of radius $r_s$) naturally arises in this approach, the expected regularization of the Riemann invariant does not occur.} From this viewpoint, Gauss-Bonnet gravity disfavours black holes as Universe-makers. Of course, this theory remains a low-energy effective description and one can still hope that a full quantum gravitational treatment changes this conclusion.

The point made here is nevertheless quite robust for two reasons. First, because
it remains true when a Maxwell term $e^{-2\Phi}F_{\mu\nu}F^{\mu\nu}$ is added to
the action, at least up to a critical value $q_{cr}$ (of the order of $\approx
0.3 r_h$ in Planck units) which was numerically determined in \cite{stas2}. When
$q<q_{cr}$, the topology of the solution is unchanged and
$R_{\mu\nu\lambda\rho}R^{\mu\nu\lambda\rho}\rightarrow \infty$ (the case
$q_{cr}<q<m\sqrt{2}$ is interesting and should be studied in this perspective).
Then, because it remains stable under higher order curvature corrections of
string gravity. As an example, the fourth order action reads as

\begin{eqnarray*}\label{err}
S & = & \frac{1}{16\pi} \int d^4 x \sqrt{-g} \biggl[ - R
+ 2 \partial_{\mu} \phi \partial^{\mu} \phi \\
& + & \lambda_2 e^{-2 \phi} L_2 + \lambda_3 e^{-4 \phi} L_3
+ \lambda_4 e^{-6 \phi} L_4 + \ldots \biggr]. \nonumber
\end{eqnarray*}
In this expression, $L_2$ denotes the second order curvature correction (Gauss-Bonnet
term), $L_3$ is the third order curvature correction,
\begin{eqnarray*}
L_3 & = &
3 R^{\mu\nu}_{\alpha\beta} R^{\alpha\beta}_{\lambda\rho}
R^{\lambda\rho}_{\mu\nu}
- 4 R^{\mu\nu}_{\alpha\beta} R_\nu^{\lambda\beta\rho}
R^\alpha_{\rho\mu\lambda} 
 +  \frac{3}{2} R R^2_{\mu\nu\alpha\beta}
+ 12 R^{\mu\nu\alpha\beta} R_{\alpha\mu} R_{\beta\nu} \\
& + & 8 R^{\mu\nu} R_{\nu\alpha} R^\alpha_\mu - 12 R R^2_{\alpha\beta}
+ \frac{1}{2} R^3 ,
\end{eqnarray*}
and $L_4$ stands for the fourth order curvature correction,
\begin{eqnarray*}
L_4 & = & \zeta (3) \Biggl[ R_{\mu\nu\rho\sigma} R^{\alpha\nu\rho\beta}
\Biggl( R^{\mu\gamma}_{\delta\beta}R_{\alpha\gamma}^{\delta\sigma} \Biggr)
\Biggr] 
 -  \xi_h \Biggl[ \frac{1}{8}
\Biggl( R_{\mu\nu\alpha\beta} R^{\mu\nu\alpha\beta} \Biggr)^2 \\
& + & \frac{1}{4} R_{\mu\nu}^{\gamma\delta} R_{\gamma\delta}^{\rho\sigma}
R_{\rho\sigma}^{\gamma\delta} R_{\gamma\delta}^{\mu\nu} 
 -  \frac{1}{2} R_{\mu\nu}^{\alpha\beta} R_{\alpha\beta}^{\rho\sigma}
R^{\mu}_{\sigma\gamma\delta} R_{\rho}^{\nu\gamma\delta}
- R_{\mu\nu}^{\alpha\beta} R_{\alpha\beta}^{\rho\nu}
R_{\rho\sigma}^{\gamma\delta} R_{\gamma\delta}^{\mu\sigma} \Biggr] \\
& - & \frac{1}{2} \xi_B \Biggl[
\Biggl( R_{\mu\nu\alpha\beta} R^{\mu\nu\alpha\beta} \Biggr)^2 
 -  10 R_{\mu\nu\alpha\beta} R^{\mu\nu\alpha\sigma}
R_{\sigma\gamma\delta\rho} R^{\beta\gamma\delta\rho} 
 -  R_{\mu\nu\alpha\beta} R^{\mu\nu\rho}_{\sigma}
R^{\beta\sigma\gamma\delta} R_{\delta\gamma\rho}^{\alpha} \Biggr] .
\end{eqnarray*}
The coefficients $c_i$ and  $\xi_i$ just depend upon the type of string theory 
considered. It can in any case (expect for SUSY-II were the situation is still 
unclear) be explicitly shown \cite{stas3} that the structure of the 
singularities is unchanged and the Riemann invariant still diverges. In a way, 
the situation is therefore even worst than in usual general relativity 
as this occurs at finite distance from the center of the black hole.\\

The main point here is that perturbative string corrections seem to disfavor 
the hypothesis of a limiting curvature which would have allowed for a junction 
layer between a Schwarzschild space and a de Sitter space within black holes. 
Otherwise stated, Gauss-Bonnet black holes are not very promising Universes 
progenitors although they seem to be free of the most dramatic problems
associated with quantum gravity effects (in particular when considering the
endpoint of the Hawking evaporation process).\\

On the other hand, it is very interesting to notice that other approaches, {\it
e.g.} taking into account nonperturbative semiclassical effects in loop quantum gravity \cite{bojo},
reach exactly the opposite conclusion ! This clearly refects our lack of
understanding of the real quantum regime in the gravitational sector.\\

The multiverse hypothesis --whether falsifiable or not\footnote{Does anyone know
of a single {\it scientific} theory which satisfies Popper's criteria ? I
personally don't. But it should be kept in mind that the history of epistemology
does {\it not} ends with Popper ! Then came Feyerabend, Bloor and Collins who
fundamentally changed our views by underlying social and esthetical
determinations in science.}--
offers a very rich framework to elaborate both on the world(s) and on our scientific
creations about the world(s). It is not even clear that it makes sense to
distinguish between both proposals. The structure of time itslef could appear as
radically modified (see, {\it e.g.}, \cite{sp} and references therein). Using the string landscape to argue in
favor of this model is probably a bit na\"{\i}ve as this still relies on an 
assumed {\it single} meta-theory. This "pluriverse" is maybe not that 
pluralistic\footnote{Seen in a modal logic framework, one could wonder what 
are the allowed domains for quantifiers.}...  As often (as usual ?) this could 
tell us much more about the schemes of our mental activities than about the 
intrinsic structure of reality\footnote{Is there such a thing ?}. But it could very 
well be that physics is now taking distances with a kind of implicit ontological 
commitment. Physics faces contingency. An exciting situation !

\end{document}